# The Dynamic Organization of Sustained Human–AI Cognition: From Construct-Level Change to Relational Structure

Zijian Ru

Le Mans Université, France

## Author Note

## Abstract

As generative artificial intelligence becomes a routine participant in writing, learning, information retrieval, analysis, decision making, and problem solving, human–AI cognition research must address not only whether AI changes psychological constructs, use intensity, or task performance, but also how human cognitive activity is organized beneath similar aggregate indicators. This article proposes a dynamic cognitive organization framework that shifts analysis from construct-level change to relational organization anchored in the person's current task-cognitive state under sustained AI participation. The framework distinguishes five relational dimensions—execution locus, cognitive governance, representational reorganization, process organization, and reachable cognitive space—and a path-specific recursive principle whereby the outcomes, costs, and experiences of one interaction may selectively reweight the future probabilities of different organizational pathways. Five sets of testable propositions follow. The same overall AI-use intensity can correspond to different cognitive organizations; similar immediate outcomes can arise from different organizations that may have different predictive value for proximal subsequent outcomes; longitudinal change in cognitive organization need not track change in overall AI-use intensity; expansion of reachable cognitive space and displacement of pre-existing or emerging human-originated pathways may coexist within the same episode; and, when different cognitive organizations recur over time, they may lead people to repeatedly practice different cognitive functions, with accumulated differences potentially corresponding to different developmental trajectories in strategies, habits, and abilities. The contribution is an analytic level and five-dimensional relational structure for describing, comparing, measuring, and testing process differences that aggregate indicators or construct-level analyses do not uniquely determine.

*Keywords:* generative artificial intelligence; human–AI cognition; dynamic cognitive organization; relational organization; cognitive governance

## 1. Introduction | When AI Becomes a Routine Participant in Cognitive Activity

Generative artificial intelligence is entering an expanding range of cognitive activities, including writing, learning, information search, analysis, decision making, and problem solving. In these settings, AI provides more than information storage, retrieval, or computational support. It can generate candidate answers, restate problems, offer explanations and counterexamples, compare alternatives, provide feedback, and adjust subsequent outputs in ongoing interaction on the basis of current input and available interaction history. Compared with earlier external tools whose behavior was relatively fixed, this generativity and responsiveness allow AI to enter multiple cognitive stages within the same task.

Existing research has described this shift from several directions. Work on cognitive offloading shows that people can transfer part of their internal processing to external resources, thereby changing the cognitive demands of a task (Risko & Gilbert, 2016). Research on the metacognitive demands of generative AI further shows that prompting, evaluating outputs, deciding whether to rely on them, and adjusting workflows all create new monitoring and control demands for users (Tankelevitch et al., 2024). A survey of knowledge workers likewise found self-reported shifts in critical-thinking activity and effort from information gathering toward verification, from problem solving toward integration of AI outputs, and from task execution toward task oversight; greater confidence in AI capability was also associated with lower self-reported critical-thinking effort (Lee et al., 2025). Educational research shows, in turn, that generative AI can enter multiple phases of self-regulation, including goal formation, resource search and integration, monitoring and evaluation, strategy support, feedback, and idea generation (Xia et al., 2026; X. Li et al., 2026).

At the same time, recent human–AI research has begun to address dynamic processes more explicitly rather than examining only static "AI effects." From a distributed-cognition perspective, Zhao and Han (2026) model human–AI interaction as a cycle comprising intent formation, representation externalization, AI generation, outcome assessment, and cognitive updating. Lu and Yan (2026) propose hybrid cognitive alignment to explain how humans and AI dynamically allocate roles, coordinate representations, and adapt as tasks change. Pedreschi et al. (2025) explicitly characterize human–AI coevolution as an ongoing feedback process in which humans and AI continually influence one another. Accordingly, this article does not claim novelty for the observations that human–AI cognition is dynamic, that humans and AI redistribute cognitive work, or that sustained interaction generates feedback.

These advances nevertheless leave a gap at a different analytic level. Recent research increasingly shows that knowing whether AI is used, how much it is used, or only the final outcome is insufficient to specify the cognitive process a person undergoes during a task. AI can participate in different cognitive and epistemic functions, and similar overall levels of use or similar final performance do not imply that those functions are allocated, controlled, and organized in the same way. Existing work can describe changes in individual constructs, covariance among multiple variables, divisions of cognitive labor, or broad human–AI interaction processes, but these descriptions do not by themselves uniquely determine who executes particular functions, who sets criteria and retains final judgment, how the problem is represented, how cognitive operations are connected, or which subsequent pathways remain reachable. Two people can therefore be similar in AI-use intensity, trust, motivation, and final performance while occupying different cognitive organizations.

This article therefore changes the analytic level rather than adding another psychological construct. The question is no longer only whether a variable rises or falls, or how much AI is used, but how different cognitive functions are configured in relation to one another when AI participates on a sustained basis. The framework proposes a relational-organization level whose direct analytic object is a human-side relational configuration anchored in the person's current task-cognitive state and how that configuration changes with interaction history. Describing this configuration may require reference to relations that cross the human–AI boundary, but the AI system's internal states are not themselves part of the analytic object. By characterizing this structure, the framework provides a structured description of process differences that aggregate indicators or construct-level analyses do not uniquely determine, making those differences available for systematic study, comparison, measurement, and testing.

Two questions must be distinguished here: what the direct analytic object is, and which relations must be referenced in order to describe it. A human-side focus does not imply that every relevant relation must remain entirely within the human. AI may execute a cognitive operation, supply a problem frame or evaluative criterion, introduce candidate explanations, or redirect an unfolding activity pathway; all of these relations may cross the human–AI boundary. What the framework tracks is how such participation changes the configuration of the person's proximal goals, task representation, specific cognitive operations, monitoring and control, and process pathways, and how that organization changes with interaction history. Human-side therefore does not mean human-isolated: whether a step is performed independently by the person, executed by AI, mediated through AI, or triggered by AI, the analysis ultimately concerns how these cross-boundary forms of participation change the relational organization of human-side cognitive activity.

# 2. Theoretical Foundations | From Cognitive Components to Dynamic Relational Organization

## 2.1 The Dynamic Unit: Why Relational Organization Can Be an Object of Analysis

The framework does not assume that “dynamic cognitive organization” is a phenomenon unique to the era of generative AI. In introducing the notion of the dynamic unit, Mandelblit and Zachar (1998) argued that the unity of an analytic unit need not derive from intrinsic properties shared by its components; it can instead arise from dynamic patterns of relation among them. Hutchins’s (2010) account of cognitive ecology likewise emphasizes that cognitive phenomena should be understood in situated contexts and that analytic units may be defined by patterns of interdependence among elements in a system.

These theories provide a metatheoretical license rather than a ready-made human–AI taxonomy. Neither proposes the five forms of reorganization developed here, nor do they claim that cognition simply is the relational structure defined in this article. What they support is a more basic analytic move: cognitive research can treat how components are organized in relation to one another as an object of analysis rather than measuring only the isolated level of each component.

## 2.2 Boundaries of Online Activity: Self-Regulation, Metacognition, Motivation, and Affect

If every variable concerning the person were incorporated into “dynamic organization” as an additional structural dimension, the framework would become open-ended rather than principled. Efklides’s (2011) MASRL model provides an important basis for drawing the boundary of the present online analysis. The model distinguishes a Person level from a Task × Person level. The former includes relatively stable characteristics such as ability, metacognitive knowledge, self-concept, perceived control, and motivational tendencies; the latter describes metacognitive experiences, online affect, task motivation, and regulatory activity occurring within a specific task. The two levels interact but are not analytically identical.

The present framework therefore focuses primarily on online processes that actually help determine what happens next in the current task. General trust in AI, long-term goals, personality, prior ability, and accumulated AI experience can influence current organization without all becoming nodes inside the online cognitive unit. Conversely, current uncertainty, effort, interest, or emotion can enter the analysis as conditions on state transitions when they alter decisions to verify, stop, continue exploring, or delegate.

### 2.3 Cognitive Control: Why "Who Executes?" Must Be Distinguished from "Who Governs?"

Research on cognitive control further shows that cognition is not a flat collection of variables operating in parallel. Badre (2025) characterizes cognitive control in terms of selecting context-appropriate behavior, monitoring ongoing behavior, and adjusting it when necessary, while also emphasizing that the content and structure of task representations shape control processes.

This distinction is especially important for human–AI cognition. When AI performs an operation, a shift in execution does not necessarily imply a corresponding shift in control. A person can let AI retrieve information, generate content, or perform calculations while still determining the problem, evidential criteria, next steps, stopping point, and final conclusion. Conversely, a person may appear to perform many operations personally while in practice accepting the problem framing, evaluative criteria, and action direction supplied by AI.

The framework therefore uses cognitive governance as a working relational analytic concept for describing how functions related to cognitive control are allocated between the human and AI. It is neither a new psychological mechanism parallel to cognitive control nor an established standardized construct. Its purpose is simply to answer a relational question: in the current human–AI activity, who sets goals and criteria, decides the next step, determines stopping conditions, and assumes final epistemic judgment?

**2.4 Representation: AI Can Change More Than the Answer**

Cognitive control depends on what is being controlled. Problem-solving theory has long emphasized that changing a problem's mental representation changes the operations that become available and the subsequent search space. Ohlsson (1984), for example, describes restructuring as a change in mental representation that alters the availability of problem-solving operations. AI therefore cannot be understood only as doing more work on a problem whose representation remains fixed.

Generative AI can provide new external representations through summarization, classification, naming, analogy, restatement, and reframing. Once these representations enter subsequent human processing, they may alter the current task model. An ambiguous, underspecified internal understanding may be unfolded into several relations that had not previously been explicit, or it may be prematurely narrowed into a highly fluent and clear but more restrictive frame. The former may open new entry points for reasoning; the latter may cause weakly articulated cues to drop out of the current working representation. The claim here concerns possible effects of AI outputs on human representation, not a specialized algorithm inside the model that "compresses human cognition." Representational change must therefore be distinguished from execution delegation: the former changes how the person currently represents the problem cognitively, whereas the latter changes who performs a cognitive operation within that problem.

### 2.5 Historical Dependence: Why One Change Can Affect the Next

Sustained human–AI cognition also requires bringing time back into the model. A dynamic approach does not treat each cognitive episode as an independent event that starts from zero. The current state is shaped by prior activity, and the outcomes of current behavior become conditions for later states. The framework applies this general principle to sustained AI use without claiming that the feedback loop itself is novel; human–AI coevolution already explicitly theorizes cycles of continuing mutual influence between humans and AI (Pedreschi et al., 2025).

In human–AI settings, a correct answer, a serious error, a clear saving in time, an expensive verification episode, or an important AI-inspired discovery may each change the probability that different pathways will be selected in a subsequent similar task. Crucially, such updating need not be fully captured by changes in overall amount of AI use. After a serious error, a person may still use AI extensively yet shift from "AI answer → direct acceptance" to "AI answer → source verification → independent comparison → final human judgment." What has changed is cognitive organization; whether or not overall amount of use changes at the same time, it cannot fully represent this structural change. The path-specific updating principle is specified in Section 3.8.

### 2.6 The Theoretical Problem Left by Generative AI

External cognitive resources, distributed cognition, cognitive offloading, self-regulation, and cognitive control all predate generative AI. The important theoretical condition introduced by generative AI is not that a tool influences cognition for the first time, but that generativity, responsiveness, and the capacity to intervene across cognitive stages can coexist within the same general-purpose system and participate continuously in a task. Such a system can execute operations, restate problems, propose candidates, provide standards, participate in verification, and continuously generate new cognitive material on the basis of current input and available interaction history.

The question therefore is no longer only "which cognition is offloaded?" but how AI, when it repeatedly enters the cognitive chain, changes the organization of human-side cognitive activity. The next section introduces five relational dimensions for characterizing the structural reorganizations that may occur.

# 3. A Framework for the Dynamic Organization of Sustained Human–AI Cognition

## 3.1 Analytic Object: Online Cognitive States and Relational Reorganization

The core analytic unit is not a "five-dimensional score vector." Within a theoretically meaningful short time window, online cognitive organization is defined as a temporary relational configuration anchored in the person's current task-cognitive state across five relational dimensions. The person's proximal goals, current task representation, and cognitive operations that the person is currently performing or monitoring constitute the content of the human-side online state; the five relational dimensions characterize structural relations between that state, task-relevant cognitive operations, and AI participation: who executes which operations, who retains effective decision authority, how the current representation is organized, how operations are connected, and which subsequent pathways remain reachable from the current state. Operations executed by AI may enter this description as one endpoint of a relation crossing the human–AI boundary, but the AI system's internal states are not themselves part of the analytic object. Trust, uncertainty, interest, emotion, effort, and resource states can influence this configuration and its next transition without thereby becoming independent structural dimensions.

The five dimensions derive from decomposing a structural question: when AI enters ongoing cognitive activity, what kinds of relations can change? Execution locus concerns the relation between a cognitive operation and its executor. Cognitive governance concerns the relation between control-related functions and effective authority over goals, criteria, next steps, stopping conditions, and final judgment. Representational reorganization concerns relations among elements within the person's current task representation and how those elements are organized. Process organization concerns relations among cognitive operations, including their ordering, connection, insertion, bypass, collapse, or recurrence. Reachable cognitive space concerns the relation between the current cognitive state and potential subsequent states that still have a chance to enter further thought. These are therefore not five homogeneous psychological variables, but five kinds of structural relation used to describe the organization of cognitive activity.

Why retain these five relational dimensions rather than adding goals, trust, motivation, emotion, or other constructs as sixth and seventh dimensions? The framework decomposes relational organization rather than enumerating every psychological variable that may matter. A candidate dimension warrants independent structural status only if it answers a distinct organizational question and can, in theory, change while other dimensions remain relatively stable. The content of a proximal goal belongs to the current online state, whereas who determines that goal belongs to cognitive governance. Trust, motivation, interest, emotion, ability, and other constructs may influence, constrain, or result from organizational transitions without thereby becoming forms of relational reorganization themselves. The five dimensions are therefore proposed as a provisionally nonredundant and parsimonious analytic set for the present theoretical problem, not as an ontological claim that human–AI cognition can only ever have five dimensions. If future empirical work identifies a relational change that cannot be represented by this set and that adds independent explanatory value, the framework should be split, merged, supplemented, or otherwise revised.

### 3.2 Execution Locus: Who Performs This Step?

The first dimension is execution locus: who primarily carries out specific cognitive operations such as retrieval, generation, comparison, calculation, expression, summarization, and revision. Research on cognitive offloading already shows that people use external resources in ways that alter the amount and kind of internal processing required by a task (Risko & Gilbert, 2016). Generative AI expands the range of operations that can be externally performed, but existing evidence does not justify equating all forms of offloading with general cognitive decline. A review focused on retrospective information shows that external storage can improve current task performance while imposing costs on internal memory under particular conditions; these conclusions should remain bounded to the relevant tasks and mechanisms (Richmond & Taylor, 2025). A recent Science & Society discussion of AI-related cognitive offloading likewise suggests that repeatedly handing certain activities to AI may impede the acquisition or maintenance of specific skills, while emphasizing that the risk depends on how AI is used and should not be generalized to a broad decline in basic cognitive ability (Cash et al., 2026).

Execution locus is therefore not synonymous with “amount of AI use.” Two people can invoke AI ten times while delegating very different operations—for example, one may delegate language polishing and the other the generation of a core argument—and the cognitive operations they perform autonomously may differ substantially.

### 3.3 Cognitive Governance: Who Steers?

The second dimension is cognitive governance: who actually determines the current goal, evaluative criteria, next step, stopping condition, acceptance or rejection, and final judgment.

Metacognitive research already treats prompting, evaluation, reliance, and workflow optimization as monitoring-and-control problems in human–AI interaction (Tankelevitch et al., 2024). Survey work with knowledge workers likewise reports self-perceived shifts in critical-thinking activity and effort toward verification, integration, and task oversight (Lee et al., 2025). More directly, Zhu et al. (2026), in distinguishing dependent from autonomous offloading, theorize cognitive agency transfer and describe the transfer of "cognitive governance" to show that execution delegation and the leadership of thinking are not the same. Their three-wave survey, however, measures self-reported perceived cognitive outcomes, and the authors explicitly limit the evidence to associations rather than objective ability change or causal proof. The present framework therefore does not claim the term cognitive governance as its invention. Instead, it uses the term as a working relational dimension for systematically describing how control-related functions are configured between humans and AI.

Execution delegation and governance delegation must therefore be separated. AI may perform a large share of operations while the human retains governance; it may also acquire greater governance influence through problem framing, evaluative criteria, or stopping recommendations even when the share of operations executed by AI changes little. The local control required to carry out a specific operation is not equivalent to higher-order governance over what the current activity is trying to achieve, how success is judged, what happens next, when to stop, and who assumes final judgment. Collapsing execution and governance into a single broad measure of "reliance" would erase this structural difference.

### 3.4 Representational Reorganization: What Does the Person Now Take the Problem to Be?

The third dimension is representational reorganization. AI-generated summaries, classifications, labels, analogies, and restatements can provide new external representations for the current task and may change the person's cognitive representation of that task, such that the problem subsequently confronted is no longer cognitively identical to the one encountered before AI intervention. Representational reorganization may expand or compress. AI may unfold a vague intuition into several distinguishable relations, or it may prematurely narrow an ambiguous and still-unstable understanding into a clear frame.

This dimension differs from reachable cognitive space because the two refer to different structural questions. Representational reorganization concerns the structure of the person's current task representation—what the problem is presently taken to be and how its elements are related. Reachable cognitive space concerns transition possibilities from that state—which candidate answers, counterexamples, hypotheses, or pathways still have a chance to enter subsequent thought. A representation can change without adding new candidate answers; conversely, the representation can remain largely stable while AI introduces several previously unconsidered options. Classical problem-solving theory proposes that representational change can alter the operations and search space subsequently available (Ohlsson, 1984). Precisely because one can causally influence the other, they should not be treated as the same relation.

### 3.5 Process Organization: How Does Cognitive Activity Unfold?

The fourth dimension is process organization: how cognitive operations are connected, ordered, inserted, bypassed, collapsed, or arranged into loops. Process organization is distinct from execution locus. The sequence "retrieve → compare → judge" may remain intact while only retrieval shifts from the person to AI. Alternatively, all key operations may remain human-executed while AI changes the pathway to "form a hypothesis → solicit AI critique → recheck sources → compare → restructure." These examples simply illustrate that pathway relations can change; they do not imply that human–AI tasks generally follow any one fixed sequence.

Recent work on “procedural collapse” in LLM-assisted writing provides a clear domain-specific example of this risk. A complete AI output can compress a process that would otherwise involve stepwise generation and revision into a structure of “receive a complete product first, then perform holistic evaluation,” thereby changing the cognitive work that actually takes place in writing (J. Kim & Mei, 2026). This work closely neighbors the present framework, but the present account does not generalize one writing-specific structure into a universal human–AI mechanism. Instead, it treats the connection pattern among cognitive operations as a relational dimension that can be observed across tasks.

### 3.6 Reachable Cognitive Space: What Still Has a Chance to Enter Subsequent Thought?

The fifth dimension is reachable cognitive space: under the current organization, which candidate problems, hypotheses, counterexamples, concepts, solutions, or pathways have a chance to enter subsequent thought. It is not a mature, standardized psychological construct, but a working analytic concept used here to examine human–AI exploration. Its concern is not simply how many ideas are ultimately generated, but which directions gain an opportunity to be further developed and expressed.

Generative AI creates a dual possibility in this respect. It can offer candidates that a person would have difficulty producing independently or help loosen existing fixation, thereby opening new directions. In experiments on design work, Hou et al. (2025) found that GenAI significantly improved creative performance during ideation, with video analysis indicating that the effect was associated with reduced cognitive fixation and greater divergent thinking. Conversely, early exposure to complete, fluent, or highly concrete AI content may become a strong anchor. In a visual ideation experiment, Wadinambiarachchi et al. (2024) found that AI image support increased design fixation, reduced the number of ideas, and lowered diversity and originality, while also documenting fixation displacement from original examples toward AI images. Research on creative writing found that AI-generated ideas could increase the average creativity of individual outputs while making different participants' outputs more similar to one another (Doshi & Hauser, 2024). A subsequent analysis of existing brainstorming data likewise reported lower group-level idea diversity in the ChatGPT condition (Meincke et al., 2025). These studies separately support the possibility of expansion/defixation and fixation/convergence, but they do not directly demonstrate that expansion and displacement of a pathway occur simultaneously within the same individual during the same cognitive episode. That stronger possibility is introduced below in P4 as a testable prediction rather than stated as an established empirical fact.

### 3.7 The Five Dimensions Are Distinguishable but Can Be Dynamically Coupled

The five forms of reorganization are analytically distinguishable but can be dynamically coupled in cognitive activity. A change in one structure may increase the probability of change in another without implying that the two are the same construct.

Consider a hypothetical coupled sequence. AI restates a research question, first changing the person's problem representation. The new representation makes previously invisible options reachable. The person decides to explore one new direction, changing a governance decision. The new exploration then requires insertion of search and comparison operations, altering process organization. Retrieval may finally be delegated to AI, changing execution locus as well. The example only illustrates that the five forms of change can trigger one another along the same activity chain; it does not predict that they will always occur as a complete sequence.

Now consider separability. The original process "retrieve → compare → judge" may remain unchanged while only the agent performing retrieval shifts from the person to AI;

execution locus changes while process organization and final governance remain stable. Process organization can also change while the set of operations, their executors, and governance allocation remain the same. For example, in two interactions the person may perform generation, verification, comparison, and final judgment while AI performs a single critique; one pathway may be "human generation → AI critique → human verification → human comparison → human judgment," whereas another is "human generation → human verification → AI critique → human comparison → human judgment." Execution locus and governance allocation are the same, while only the ordering and connection among operations differ. Conversely, the same surface sequence—"AI generates → person reads → submits"—may involve a person who actively judges according to independently established criteria or one who simply accepts the criteria supplied by AI. The process can therefore appear similar while governance differs. The theoretical possibility of changing A while leaving B relatively stable is the minimum requirement for treating the five dimensions as meaningfully distinguishable.

### 3.8 The Recursive Dynamic Principle and Multiple Time Scales

The recursive principle is not a sixth structural dimension. The proposed dynamic principle is: current cognitive organization → human action and AI response → outcomes, costs, and experiences → selective reweighting of the future probabilities of different organizational pathways → next cognitive organization. Here, an "organizational pathway" is a distinguishable transition pattern constituted across multiple relational dimensions, not a single isolated act of verification, delegation, or exploration. The claim is not merely that past experience affects future behavior, that successful delegation can reinforce later delegation, or that generic feedback loops exist. More specifically, updating may be pathway-specific within a multidimensional cognitive organization: an error may increase the future probability of configurations involving

"AI execution + inserted verification + retained human final judgment"; a substantial time saving may increase the probability of "greater execution delegation + retained human governance"; and an unexpected discovery may increase the probability of "AI-introduced candidates + expansion of reachable space + continued human exploration," even when aggregate AI-use intensity remains relatively stable. The framework does not posit a novel psychological learning mechanism for this reweighting and remains agnostic about whether it is implemented through reinforcement, belief updating, strategy learning, or other adaptive processes; established learning and adaptation mechanisms may provide its psychological implementation. Its proposed increment concerns the target and granularity of updating: whether outcomes, costs, and experiences selectively alter future probabilities over distinguishable multidimensional organizational pathways in ways not fully captured by aggregate AI-use intensity, dependence, trust, or a general delegation tendency. Accordingly, if such experiences predict only broad tendencies and provide no additional prediction of pathway-specific organizational transitions, the recursive principle would not be supported. Recent work has theorized successful delegation as a self-reinforcing process and linked GenAI-use intensity, dependence, and perceived skill atrophy in a two-wave survey (Subramanya et al., 2026). The framework therefore does not claim novelty for self-reinforcement itself.

These changes must be distinguished across time scales. At a seconds-to-minutes scale, the object of study is the online state and structural transitions within a task. Across a task or several weeks, some pathways may develop into relatively stable strategies or recurring patterns. Only over months or longer is it appropriate to investigate stable role divisions, changes in higher-order goal structures, ability development, and cognitive habits. Persistent shifts in pathway probabilities may be one source of the redistribution of cognitive practice opportunities discussed below, but they are not a necessary condition for P5. The recursive principle and P5 address questions at different levels: the former tests whether outcomes, costs, and experiences in interaction selectively alter the subsequent probability of different organizational pathways; the latter begins from the fact that a particular cognitive organization recurs across repeated or functionally comparable tasks and tests whether that recurrence systematically changes the practice and feedback opportunities available to different human-side cognitive functions. P5 therefore does not require the recurrence of a particular organization to be caused by the recursive principle; as long as an organization recurs stably, whether practice opportunities are systematically redistributed as a result can be tested independently.

At the level of time scale, adjustments to proximal goals are part of the current online organization, whereas whether sustained AI use alters a person's higher-order goal system is a potential outcome at a slower time scale and should not be repackaged as a sixth online structural dimension.

## 4. Five Sets of Testable Propositions

The five sets of propositions form a progressively structured theoretical argument. P1 and P2 begin with two common aggregate indicators—overall AI-use intensity and immediate outcomes—to test the discriminative value of relational organization as a distinct analytic level: neither the same amount of use nor a similar immediate result automatically determines the cognitive organization beneath it. P3 extends this issue longitudinally by asking whether changes in aggregate use intensity can substitute for observing change in cognitive organization itself. P4 and P5 then advance substantive predictions generated by the relational framework: P4 concerns the coexistence of expansion of cognitive space and displacement of human-originated pathways within the same cognitive episode, whereas P5 connects sustained organizational differences to cognitive practice opportunities and possible developmental trajectories.

**P1 | The Same Overall AI-Use Intensity Can Correspond to Different Cognitive Organizations**

Frequency, duration, and number of AI interactions describe use intensity but do not uniquely determine a person's cognitive organization. The same overall amount of use can underlie different pathways, such as "AI generates → direct acceptance," "AI generates → cross-checking → human judgment," or "human generates first → AI challenges → human restructures." Zhu et al.'s (2026) three-wave survey shows that dependent and autonomous cognitive offloading are distinguishable forms of AI engagement with different agency- and motivation-related correlates. The evidence is correlational and centers primarily on perceived outcomes, but it directly supports the weaker premise that usage mode cannot be reduced to frequency. Gordetzki et al. (2026) further use a fine-grained framework of human–AI agency configurations to explain differences in creativity and effort across AI representation conditions, while process research has begun to identify distinct AI-supported self-regulated-learning event sequences from behavioral logs (Tao et al., 2026). On this basis, the framework proposes that the same amount of AI use can conceal systematically different configurations of execution locus, cognitive governance, representational reorganization, process organization, and reachable cognitive space.

The point of P1 is not to establish the already-known fact that people use AI differently. It is a discriminative and incremental prediction: use intensity, despite being a common metric, may systematically mask differences in cognitive organization, and measuring those differences should explain or predict variance that aggregate use does not. Even when overall amount of AI use is equal or comparable, and even when conventional usage characteristics are similar, different users may develop different configurations of execution locus, cognitive governance, representational reorganization, process organization, and reachable cognitive space. These configurations may in turn correspond to differences in verification, exploration, transfer, practice opportunities, and subsequent trajectories. A direct test of P1 should therefore compare models with and without relational-organization measures and ask whether the former provide incremental explanatory or predictive value after controlling for amount of use, task, individual ability, and conventional usage characteristics. If they do not, the incremental explanatory or predictive value claimed by the framework at the level of P1 would be weakened.

### P2 | Immediate Outcomes Cannot Exhaust the Cognitive Process or Its Subsequent Trajectory

Immediate performance under GenAI support cannot automatically be equated with learning, nor does it exhaust the other consequences that may arise from a human–AI cognitive episode. Yan et al. (2025) explicitly distinguish performance gains from learning and emphasize that high-quality learning still requires appropriate cognitive and metacognitive processing. A preregistered randomized experiment by Kreijkes et al. (2026; N = 405, students aged 14–15) provides a concrete but bounded educational example. Note-taking alone and “LLM + notes” both outperformed LLM-only use on comprehension and retention three days later. Yet in the comparison between LLM use and note-taking alone, students preferred the LLM, regarded it as more helpful, and reported greater activity enjoyment, while task interest did not differ significantly. Open-ended responses and prompt behavior also showed that some students used the LLM to ask about background, explanations, and questions that extended beyond the source text while remaining relevant to the topic. These results show that learning performance, subjective experience, and exploratory behavior can move in different directions, but they do not establish that such process or experience differences produce longer-term knowledge expansion.

The framework does not present “performance ≠ learning” as a new finding. Existing research already shows that similar immediate performance does not imply the same cognitive process or learning outcome. P2 advances this point by using the relational-organization framework to characterize otherwise broad “process differences” in structural terms. Similar current outputs can arise from different combinations of autonomous generation, verification, revision, exploration, and control; the five-dimensional framework asks not merely whether these activities occur, but how they are organized in human–AI interaction: who performs which operations, who retains goals, criteria, and final judgment, how the current task representation is reorganized, how the activities are connected into a process, and which subsequent cognitive pathways remain reachable. Immediate outcomes therefore cannot uniquely identify the cognitive organization that produced them. If these organizational differences are theoretically meaningful, they should also provide additional prediction of more proximal subsequent outcomes, such as later independent performance, transfer, or continued exploration. If no reliable organizational differences can be distinguished when immediate outcomes are matched, or if established organizational differences provide no additional proximal predictive value, the corresponding claim of P2 would be weakened. How sustained organization may contribute to strategy, habit, and ability development through the long-term accumulation of differentiated practice opportunities is reserved for P5.

**P3 | Changes in AI-Use Intensity Need Not Track Changes in Cognitive Organization**

Among people engaged in sustained AI use, adaptation need not appear simply as “using more” or “using less.” A randomized vignette experiment by Yu et al. (2026) provides an important but bounded indication: a safety interface significantly increased verification intention, whereas reliance intention did not change significantly; the behavioral proxy of expanding source information showed only a trend and did not reach the conventional significance threshold. The study did not manipulate AI errors and did not establish either sustained real-world verification behavior or stability in long-term amount of use, so it cannot directly test the present proposition.

What the Yu et al. (2026) study can support is a weaker premise: verification intention and reliance intention need not move in synchrony as simple inverse indicators. Building on that dissociation, P3 makes a within-person longitudinal prediction: changes in cognitive organization that arise with interaction history need not move in synchrony with changes in overall AI-use intensity. After consequential feedback about AI reliability, for example, a user may reconfigure verification, cognitive governance, or process pathways without a corresponding change in the same direction, at the same time, or of the same magnitude in overall frequency, duration, or number of AI interactions. P3 therefore concerns whether organizational change can be relatively decoupled from change in aggregate use indicators, not merely whether cognitive organization changes at all. Tracking only whether a person uses AI "more" or "less" may fail to capture such adaptation. If changes in cognitive organization always move in synchrony with overall AI-use intensity and the latter is sufficient to explain the former, P3's independent claim would be weakened.

**P4 | Expansion of Cognitive Space and Displacement of Human-Originated Pathways Can Occur Together**

Existing studies separately show that GenAI can reduce some forms of cognitive fixation, promote divergence, and improve creative performance during ideation (Hou et al., 2025), while also providing individuals with outputs that receive higher creativity ratings. At the same time, AI assistance can increase design fixation, reduce group-level diversity, or, under particular representation conditions, be theorized as constraining subsequent human agency (Doshi & Hauser, 2024; Wadinambiarachchi et al., 2024; Meincke et al., 2025; Gordetzki et al., 2026). Because these findings come from different tasks and analytic levels, they cannot be combined into an established claim that AI has suppressed a particular latent idea pathway within an individual.

The additional step proposed here is therefore retained as a theoretical prediction: within the same person and the same cognitive episode, AI may increase newly reachable cognitive states while reducing the probability that some pre-existing or emerging human-originated pathways will continue to develop or be expressed. The framework provisionally uses expansion–displacement to describe this co-occurrence without presenting it as an already established construct. The prediction does not require researchers to read completely unexpressed “latent thoughts”; it concerns only pathways for which there is evidence of existence or emergence before AI intervention.

Future tests must therefore first establish pre-intervention evidence for candidate pathways—for example through preliminary notes, sketches, problem formulations, think-aloud segments, hypotheses, or other process traces—rather than inferring after the fact that an invisible idea must once have existed. They should then test whether AI intervention changes the probability that these emerging pathways continue to develop or be expressed while also introducing newly reachable candidates. To attribute displacement to AI intervention, an appropriate control or comparable baseline is also required to distinguish AI-induced changes in pathway persistence from the natural attrition that would occur during cognition anyway. If AI merely adds new candidates without systematically altering the persistence probability of pre-existing or emerging pathways supported by prior evidence, the co-occurrence of expansion and displacement proposed in P4 would lack support.

**P5 | Sustained Cognitive Organization First Redistributes Practice Opportunities and May Subsequently Produce Long-Term Cognitive Differentiation**

For longer-term outcomes, the framework distinguishes two levels. The first is a more proximal prediction about practice opportunities. The FLoRA/HHAIRL work explicitly proposes that AI development may either diminish or enhance learners' opportunities to practice their own regulatory skills (X. Li et al., 2026); Xia et al.'s (2026) systematic review of 73 studies further shows that GenAI can enter multiple self-regulated-learning activities. Tao et al. (2026), in a specific English-speaking task conducted in a GenAI-supported immersive VR environment, use fine-grained behavioral logs to identify strategic and reactive self-regulated-learning process patterns and report that these patterns are associated with different perceived gains in self-regulation. Chen's (2026) eight-week quasi-experiment further distinguishes performance under AI support from near-transfer performance without AI: deeper idea/reasoning offloading in the open-collaboration condition was associated with lower independent higher-order performance at Week 8. Because conditions were assigned to only six intact classes and the bounded-support condition combined limits on delegation with compulsory reflection, these results cannot establish definitive long-term causal relations. The weaker premise they support is that different forms of AI participation can change which cognitive and regulatory activities people actually experience during a task.

P5a: If different cognitive organizations recur across repeated or functionally comparable tasks, they should form different distributions of cognitive practice opportunities. "Practice opportunities" here are not a relabeling of single-episode cognitive organization; they refer to the cross-task accumulation of which cognitive operations the person actually performs repeatedly, which governance functions they assume, and the associated effort and feedback exposure. For example, one organization may reduce autonomous initial generation while increasing verification, comparison, and integration. Different organizations do not by definition necessarily produce different cumulative practice distributions; if task demands and available opportunities are comparable or controlled and different organizations still do not show systematic differences on these independently accumulated indicators, P5a would lack support.

The theoretical increment of P5a is not to propose a new practice effect or skill-acquisition mechanism, but to raise a more upstream question: in sustained human–AI cognition, does relational organization systematically determine which cognitive functions people repeatedly practice? Existing theories of learning and skill acquisition can explain why differentiated practice may produce developmental consequences; the present framework tests how such differentiated practice opportunities arise from different cognitive organizations.

The second level concerns long-term development. AI offloading may affect the acquisition or maintenance of particular skills (Cash et al., 2026), and conceptual work on the division of cognitive labor and metacognitive oversight has proposed that different human–AI configurations may correspond to different capability trajectories (T. W. Kim et al., 2026). Existing evidence nevertheless remains insufficient to establish that a particular distribution of practice opportunities, sustained over time, necessarily produces a stable configuration of ability.

P5b: If different distributions of cognitive practice opportunities remain stable over time, they may further correspond to different developmental trajectories in strategies, habits, and abilities. This prediction does not require overall cognitive ability to uniformly improve or decline; different functions may develop in different directions. Whether such differentiation actually emerges requires separate longitudinal measurement of strategies, habits, and abilities, together with tests of whether these developmental changes correspond to long-term differences in the preceding distributions of cognitive practice opportunities. If stable differences in practice opportunities do not correspond to replicable differences in strategies, habits, or abilities, P5b would lack support.

## 5. Methodological Implications | How to Observe Dynamic Cognitive Organization

### 5.1 From Surface Behavior to Process Structure—Without Treating Behavior as the Internal State

If the framework directly analyzes relational organization, methods cannot stop at recording whether AI was used, how many times it was invoked, or whether an answer was accepted. The same acceptance behavior can arise from high trust, time pressure, acceptance after verification, agreement with one's own judgment, low task value, or lack of independent judgment. Conversely, frequent verification can reflect low trust, high risk, professional norms, or a stable metacognitive habit.

The methodological challenge is therefore to avoid a common inference error: observing behavior and claiming that the internal cognitive state has been directly measured. Interaction logs, prompts, AI responses, clicks, document versions, and timestamps can reliably describe observable processes. Think-aloud data, brief state reports, confidence judgments, and targeted interviews can provide additional grounds for inferring functional meaning. Internal states, however, still require explicit operationalization and converging evidence from multiple sources; they cannot be read directly from surface behavior or aggregate indicators.

## 5.2 The Five Relational Dimensions Require Different Operationalizations

Execution locus can be operationalized through task decomposition and version histories that identify who primarily performed particular operations. Cognitive governance requires evidence about who established goals and criteria, who decided to continue or stop, who initiated verification, and who retained final judgment; number of AI interactions is not a direct proxy for governance. Representational reorganization requires externalizing task understanding before and after AI intervention through, for example, problem definitions, concept maps, preliminary explanations, or hypotheses. Process organization requires timestamps and sequence data to reconstruct how cognitive steps were connected. Reachable cognitive space requires externalizing candidate problems, hypotheses, counterexamples, solutions, or other potential subsequent pathways at key time points and combining those reports with process traces to infer which directions still had a chance to enter subsequent thought. Tests focused specifically on P4 must also record pre-existing or emerging human-originated pathways before AI intervention so that "new candidates were added" can be distinguished from "existing pathways became less likely to persist."

A single study need not measure all five relational dimensions. The theoretical requirement is to collect the minimum evidence sufficient to discriminate the target proposition from plausible alternative explanations, rather than placing every psychological variable, log measure, and physiological indicator into the same design. Online probes used to externalize candidate pathways may themselves alter the unfolding cognitive process. Measurement reactivity can be reduced by limiting the number of probes, avoiding prompts that supply candidate content, and combining online process traces with post-task stimulated recall. Zhong and Zhu (2026) provide a feasible example: they relied primarily on screen recording and eye-tracking during the task and then used participants' own interaction videos as recall cues in post-task interviews, combining observable trajectories with retrospective process interpretation.

**5.3 The Design Must Match the Time Scale of the Claimed Change**

Delegating execution in one task cannot establish skill decline, and a single act of verification cannot establish improved verification ability. Online reorganization, strategy stabilization, and long-term development require different designs. At the seconds-to-minutes scale, high-temporal-resolution logs, state reports, and experimental manipulations can examine transitions. Across a task or several weeks, studies can examine whether particular pathways develop into relatively stable strategies or recurring patterns. Claims about change over months or years require repeated longitudinal measurement together with no-AI transfer tasks, independent generation tasks, error-detection tasks, memory and comprehension measures, or problem-finding tasks.

This is also why P5a and P5b must remain separate. Current process data can tell us, relatively directly, what people repeatedly practice; they cannot substitute for long-term measures of ability. Tests of P5a should not treat repeated organizational codes themselves as the practice-opportunity outcome; they should separately accumulate across tasks the actual practice frequency of specific cognitive operations or governance functions, the proportion undertaken independently, effort investment, and feedback exposure, and test whether these cumulative indicators systematically shift with recurring organizations. Practice-opportunity indicators used to test P5a should be operationally distinguished from the coding criteria used to identify cognitive organization, avoiding definitional overlap or any specification that makes differences in practice opportunities follow directly from the organizational classification criteria. Tests of P5b should additionally separate cross-task cumulative practice-opportunity indicators from subsequent indicators of strategies, habits, or abilities, testing whether the former correspond to longitudinal change in the latter rather than using the same set of process codes to serve simultaneously as practice opportunities and developmental outcomes. Tests of the recursive principle additionally require repeated or continuous coding of organizational configurations and their transitions across ongoing interaction or repeated/functionally comparable tasks, followed by tests of whether particular outcomes, costs, or experiences selectively change the subsequent probability of different organizational pathways rather than merely predicting change in overall use intensity or broad tendencies such as dependence or trust.

### 5.4 The Most Informative Comparison Is Not Always “AI vs. No AI”

AI-versus-no-AI comparisons continue to answer important questions. For post-adoption or sustained-use settings, however, other comparisons may map more directly onto the present framework: compare different divisions of cognitive labor, governance structures, process pathways, and verification organizations when AI availability and overall use intensity are comparable or controlled; or track changes in the same participant’s pathways over weeks or months.

Such designs do not abandon causal inference. They distinguish the effect of AI being present from the relational organization and change of human-side cognitive activity once AI has already become a sustained participant. A theoretical proposition may examine the idealized case in which organization differs despite identical overall amount of use. Empirical studies can approximate this condition more realistically through matching, statistical adjustment, or experimental constraints that create comparable amounts of use.

## 6. Theoretical Contribution, Nearest Neighbors, and Boundaries

### 6.1 Relation to Dynamic Frameworks of Human–AI Cognition

Zhao and Han (2026) explicitly model human–AI interaction as a jointly constituted human–AI cognitive system comprising four subsystems: a user cognitive subsystem, an interaction interface subsystem, an AI subsystem, and an external representation subsystem. Their dynamic mechanism is organized into five temporal phases: intent envisioning, representation externalization, generative reasoning, outcome assessment, and cognitive update. Lu and Yan's (2026) hybrid cognitive alignment likewise addresses dynamic task and role allocation together with representational coordination. These accounts already cover broad claims about dynamic processes, mutual adaptation, role change, representational coordination, and post-adoption cognitive organization. The contribution of the present framework therefore cannot rest on claiming these phenomena as novel.

Zhao and Han also explicitly discuss movement of the processing locus toward the AI subsystem, users' retention or relinquishment of core control in planning and evaluation, task formation, verification, cognitive updating, and longer-term cognitive capability evolution. Broad claims such as "AI reallocates cognitive labor," "execution can shift while control remains human," or "different interaction organizations can produce different long-term cognitive outcomes" therefore cannot serve as novelty claims here either. The two frameworks nevertheless have different structures. Zhao and Han center a coupled system, representational flow, five temporal phases, three types of gap, and cycle completion/truncation. The present framework instead centers the human side and treats execution locus, cognitive governance, representational reorganization, process organization, and reachable cognitive space as five relational dimensions that can vary across phases. In particular, Zhao and Han treat retention or relinquishment of control as one important condition differentiating outcomes of the interaction cycle, whereas the present framework treats governance allocation as a relational dimension that can be analyzed separately from execution locus, representational reorganization, process organization, and reachable cognitive space. It further proposes a recursive prediction in which distinguishable multidimensional relational-organization pathways are the target of updating, together with predictions about differentiated distributions of practice opportunities.

Several other 2026 contributions further narrow the space for broad novelty claims. Qiang's (2026) SSRN position paper treats cognitive organization and collaborative organization as analytically separable dual states and discusses how they may diverge, lag, or transform into one another during long-term human–GenAI collaboration. "Cognitive organization," "relational organization," or longitudinal nonequivalence is therefore not itself an original contribution of the present framework. The distinction developed here is narrower: the analytic focus returns to the human side, five specific forms of relational change are distinguished, and a recursive prediction targeting distinguishable multidimensional relational-organization pathways is added alongside practice-opportunity predictions.

P. Li et al. (2026) introduce Cognitive Collaboration Dialogue (CCD) in a journal study that models exploratory dialogue as a process through which humans and AI progressively construct a shared problem understanding using trajectory planning, interaction coordination, and monitoring/regulation. The core focus of the published account is collaborative dialogue and shared problem understanding. The present framework instead focuses on five forms of relational reorganization on the human side. Thus, even though the analytic focus differs, "dynamic shared understanding" or "cognitive alignment" cannot itself be claimed as novel here.

A domain-specific near neighbor that appeared during final review further constrains broad novelty claims. In travel decision making, Y. Kim, Tussyadiah, and B. Kim (2026) explicitly ask how GenAI changes the cognitive organization of decision making and propose three scaffolding pathways: executive-load mitigation, psychological-processing balance, and decision-loop support. Their framework also discusses how AI redistributes cognitive work between people and digital systems and how it maintains continuity across iterative planning, with initial qualitative grounding from task interviews with 30 UK adults. Zhong and Zhu (2026) further analyze fine-grained cognitive events from 150 undergraduates performing programming tasks, using clustering and sequential pattern analysis to identify AI-reliant, AI-independent, and human–AI hybrid cognitive modes and to characterize both the distribution and sequential structure of cognitive events in those modes. Their study thus provides direct empirical evidence that cognitive work can form different distributions and process patterns across humans and AI, and it links those modes to cognitive load, attention allocation, and programming performance. The present framework differs not by again claiming that different forms of AI use correspond to different cognitive processes, but by using five separately analyzable relational dimensions rather than a set of empirically clustered cognitive modes, and by asking how those relational configurations undergo path-specific change with interaction history and what subsequent trajectories they may support. Accordingly, "AI changes cognitive organization," "AI redistributes cognitive work," "different AI-use patterns correspond to different cognitive processes," and "process continuity can span iterations" are not novelty claims of the present framework. Its more specific object is a human-side relational decomposition applicable across task contexts: how the five relational dimensions form configurations and shift with interaction history.

**6.2 Analytic Level: From Multiple Variables to Relational Organization**

The interactions among self-regulation, metacognition, motivation, affect, cognitive load, trust, and agency are already addressed by mature theories and a large human–AI literature. The contribution of the present framework is therefore not to add another variable to this potentially open-ended construct space, but to propose a relational-organization level distinct from both construct-level analysis and analysis of the human–AI system as a whole: relational organization on the human side, anchored in the person's current task-cognitive state. At this level, the central question is how this human-side state forms concrete configurations across execution locus, cognitive governance, representational reorganization, process organization, and reachable cognitive space, and how those configurations change with interaction history. AI may enter the description as one endpoint of relations crossing the human–AI boundary, but the AI system's internal states are not part of this analytic level.

The theoretical value of this analytic level is both discriminative and predictive. Similar levels of trust, cognitive load, AI-use intensity, or immediate performance can occur within different cognitive organizations; even knowing that "cognitive offloading occurred" does not uniquely specify which operation was delegated, whether substantive governance was retained by the human, whether the problem representation changed, whether the process was compressed or reorganized, or which subsequent pathways remained reachable. The five relational dimensions therefore provide a common structural language for distinctions that have been dispersed across different theoretical traditions. Their value does not depend on claiming that delegation, control, representational change, or dynamic adaptation is itself new. It depends on whether jointly characterizing the five dimensions can transform process differences that remain difficult to distinguish beneath similar aggregate indicators into comparable, measurable, and testable organizational configurations, and further test whether those configurations correspond to different patterns of verification, exploration, transfer, practice opportunity, and subsequent trajectory.

**6.3 Relation to Cognitive Control and Human–AI Agency Configurations**

The redistribution of cognitive control and variation in human–AI configurations already have direct theoretical neighbors. T. W. Kim et al. (2026) use a Division of Cognitive Labor × Metacognitive Oversight framework to discuss different human–AI configurations and capability trajectories. Zhu et al. (2026) theorize different forms of engagement and shifts in governance in terms of dependent/autonomous offloading and cognitive agency transfer, supported by correlational evidence from a three-wave self-report survey. Gordetzki et al. (2026) use agency configurations to explain differences in creativity and effort across AI representation conditions. The more recent AIRIS preprint frames the problem explicitly as cognitive control allocation in hybrid human–AI cognition and proposes interacting destabilization mechanisms such as delegation drift and calibration drift, locating some changes in cumulative and regulatory dynamics across repeated interactions (Kuhn et al., 2026). Fábrega's (2026) epistemic ownership/cognitive firm framework likewise treats GenAI use as an architecture for allocating cognitive work between internal and external resources and places the human capacity to reconstruct, assimilate, and defend the reasons underlying a final cognitive product at the center of cognitive governance and epistemic ownership. Mühlhoff (2026), focusing on concrete practices of LLM use, treats the degree of delegated judgment and the degree of epistemic control retained by users as analytically separable dimensions. Du and Yuan's (2026) critical-integrative review further locates the boundary between productive reliance and harmful dependence in whether the epistemic work required to form judgment is preserved or displaced, discussing verification, evaluative judgment, and responsibility through criteria such as contestability, recoverability, and transfer. Gao and Zhang's (2026) grounded theory study likewise shows that GenAI-mediated offloading in doctoral learning can coexist with different degrees of learner-controlled engagement; when learners reconstruct AI outputs into defensible understanding and research judgment, verification, reasoning, and final judgment remain with the learner. This provides recent qualitative process evidence for differences in governance and process, but does not propose the present five-dimensional relational structure. Accordingly, the claims that amount of AI use is insufficient to characterize the organization of cognitive work, that delegation–supervision–reintegration configurations differ, and that externally performed cognition may or may not remain under substantive human governance cannot be claimed as novel here. Nor are control redistribution, delegation drift, generic dynamic instability, or self-reinforcement novel contributions of the present framework.

The framework does not treat control or governance functions themselves as new theoretical objects. Within the five relational dimensions, cognitive governance characterizes the distribution of effective decision authority over goals, standards, direction, stopping conditions, and final judgment, alongside execution locus, representational reorganization, process organization, and reachable cognitive space as part of the structural description of relational organization. The recursive principle likewise does not claim feedback, drift, reinforcement, or self-reinforcement as novel, nor does it posit a novel learning mechanism. Its proposed increment concerns the target and granularity of updating: whether different outcomes, costs, and experiences selectively reweight the future probabilities of distinguishable multidimensional organizational pathways in ways not fully captured by overall AI-use intensity or other broad tendencies such as dependence, trust, or a general tendency to delegate. The framework therefore does not simply rename the facts that control can transfer or that different agency configurations exist. Its substantive test is whether jointly characterizing the five relational dimensions provides additional discriminative and predictive information beyond aggregate use indicators and existing construct-level or configuration-level descriptions.

### 6.4 Long-Term Cognitive Outcomes: Differentiation Rather Than Presumed Unidirectional Enhancement or Decline

Cognitive offloading can improve current performance while, under particular retrospective-memory conditions, carrying costs for internal memory (Richmond & Taylor, 2025). In AI settings, sustained outsourcing of certain activities may impede the acquisition or maintenance of specific skills (Cash et al., 2026). At the same time, generative AI can create new cognitive materials and task opportunities. A person may perform less autonomous retrieval but more cross-source comparison; generate less initial text but engage in more evaluation and integration; or offload routine operations while entering problems that were previously beyond independent reach.

The framework therefore does not predict that overall cognitive ability will uniformly improve or decline in a single direction. The more disciplined claim is that if different cognitive organizations recur across repeated or functionally comparable tasks, they should form different distributions of cognitive practice opportunities; if these differences remain stable and accumulate over time, they may further correspond to different developmental trajectories in strategies, habits, and abilities.

Recent work sharpens this boundary. Lin and Al-Hada (2026) propose a product–process dissociation in GenAI-integrated learning, distinguish episodic from habitual offloading, and use cognitive debt to characterize a potential accumulating risk that may continue to affect metacognitive calibration and unassisted higher-order performance after AI assistance ends. They explicitly formulate these relations as hypotheses requiring experimental and longitudinal tests rather than established facts. Subramanya et al.'s (2026) two-wave survey of knowledge workers further reports significant associations between GenAI-use intensity and dependence, and between dependence and perceived skill atrophy, while theorizing a self-reinforcing pathway of repeated delegation, dependence, and skill costs from cognitive offloading, skill decline, and habit formation. The outcome remains perceived skill atrophy rather than objective longitudinal skill measurement. Thus, "immediate output can diverge from deep learning," "habitual AI offloading may carry long-term cognitive costs," "repeated successful delegation may reinforce dependence," and "lack of practice may be associated with skill costs" cannot serve as novelty claims here. The more specific contribution is to use the five relational dimensions to characterize which cognitive operations and governance functions are repeatedly retained, reduced, or added in human-side cognitive activity, and to test whether, when a particular organization recurs across repeated or functionally comparable tasks, the long-term accumulation of differentiated practice opportunities corresponds to different trajectories in strategies, habits, and abilities.

**6.5 Scope Conditions**

The framework is designed primarily for post-adoption or sustained-use settings in which AI has become a relatively stable cognitive resource and adaptation continues after adoption. It does not assume that everyone will continue using AI, nor does it attempt to provide a universal theory spanning all stages from pre-exposure through first adoption to long-term use. For people who have never used AI, have only experimented with it, or are still deciding whether to adopt it, technology-acceptance and initial-trust theories remain more directly applicable.

The framework also does not apply equally to every activity labeled as "AI use." If AI performs only fixed, peripheral operations that have little effect on core cognitive functions, a full dynamic-organization analysis may be unnecessary. The framework is most relevant when AI repeatedly enters core cognitive chains involving generation, explanation, representation, comparison, feedback, judgment, verification, or decision making.

## 7. Discussion and Research Agenda | From "Does AI Have an Effect?" to "How Are Cognitive Processes Reorganized Under Sustained AI Participation and Changed With Experience?"

### 7.1 Stable Use Can Conceal Ongoing Adaptation

If P1 and P3 hold, longitudinal human–AI research should not treat adoption, frequency, or continuance intention as sufficient indicators of adaptation. A person may report "using AI every day" both before and after a period of adaptation while shifting from direct acceptance to systematic verification, or from independent initial generation to AI-first generation. What has changed is the relational organization of the AI-use process.

Future studies therefore need to record both use intensity and organizational structure. Otherwise, an apparently stable frequency curve may conceal substantive cognitive adaptation.

### 7.2 "Human in the Loop" Must Be Decomposed into Specific Cognitive Functions

Human-centered AI often emphasizes human oversight, human-in-the-loop arrangements, and human agency. Yet the fact that a person performs the final click does not establish that substantive governance has been retained. A person may manually execute many steps while lacking independent criteria; another may delegate extensive execution to AI while continuing to establish standards, verify evidence, and assume final judgment.

Research on agency and oversight can therefore further distinguish who executes specific operations, who sets goals and criteria, how responsibility for verification is allocated, how cognitive operations are organized, and who assumes final epistemic judgment. Preserving human agency need not mean forcing people to perform more steps personally; it may instead mean deliberately retaining the governance functions that matter for the task's goals.

### 7.3 AI Design May Shape Cognitive Organization, Not Only Efficiency

If the timing, completeness, and representational form of AI outputs can alter process organization and exploration, interface design is also a form of cognitive-organization design. A system may deliver a complete answer immediately or first require the user to externalize an initial judgment. It may provide a single high-confidence option or juxtapose competing frames, display uncertainty, require verification, or preserve the person's original branches before offering an AI-generated restructuring.

Such designs may create different cognitive practice environments even when current task performance is similar. The framework therefore proposes cognitive-organization-aware AI design as a research direction. Its goal is not to assume that "more human work is always better," but to determine, in relation to the task objective, which cognitive functions are appropriate to outsource, which should be retained, and when and in what form AI should intervene. Learning tasks, pure production tasks, and high-risk decisions are likely to require different principles of cognitive organization.

### 7.4 Long-Term Research Should Track "What People Repeatedly Do," Not Only "How Long They Use AI"

The longitudinal research demanded by P5 is not simply a correlation between time spent using AI and ability several years later, nor must the primary design always be "AI vs. no AI."

The more direct question is, within real cognitive environments where AI remains available, to repeatedly record the actual distribution of an individual's cognitive operations and governance relations across repeated or functionally comparable tasks: whether independent problem definition is consistently retained, whether initial generation gradually decreases, whether verification and integration increase, whether stable multi-AI verification pathways emerge, and how these organizations change jointly with AI capabilities and interaction experience.

At the same time, any claim that human ability itself has changed requires periodic diagnostic measurements capable of isolating human-side change. These may include brief no-AI transfer or independent-generation tasks, or comparisons conducted with standardized AI capabilities, a fixed model, or controlled assistance. The purpose is not to return to a simple "AI effect" design, but to distinguish what changes in joint human–AI task performance, what changes in the AI system's own performance, and what changes in the person's independent or transferable cognitive ability.

**7.5 Role Specialization Is a Developmental Question, Not an Established Fact**

When a particular organization recurs over long periods across repeated or functionally comparable tasks, people may come to occupy more stable human–AI cognitive roles, such as initial generator, verifier/auditor, integrator, or problem framer. Knowledge workers report shifts in cognitive activity and effort from information gathering toward verification and from execution toward oversight (Lee et al., 2025). Procedural Collapse further proposes that LLM-assisted writing can move the person's work from stepwise generation toward holistic evaluation of a complete output (J. Kim & Mei, 2026). T. W. Kim et al.'s (2026) conceptual framework discusses potential capability trajectories associated with different configurations of cognitive labor division and metacognitive oversight. None of these accounts, however, provides sufficient long-term objective ability evidence to justify claiming that "sustained AI use has already trained people into stronger auditors."

Role specialization is only one possible manifestation of P5b, not the definition of P5b itself. It should therefore remain a research question: does repeatedly occupying a particular human–AI cognitive role over long periods change the corresponding abilities, preferences, habits, and agency? "The auditor becomes stronger" is a testable hypothesis, not a conclusion that can be assumed within the theoretical example itself.

### 7.6 A More General Methodological Shift

The broader aim of the framework is not to replace existing fine-grained empirical research with a single grand theory, but to encourage researchers to distinguish three analytic questions: what changes at the construct level, how cognitive functions form relational configurations, and how those configurations shift with interaction history. Construct-level analyses and average effects remain necessary, but once AI becomes a sustained participant, the same amount of use and the same outcome may still correspond to different processes.

The framework also has explicit conditions for revision. If the five relational dimensions cannot be reliably distinguished, or if relational organization consistently provides no incremental discriminative, explanatory, or predictive value after controlling for use intensity, individual ability, and task characteristics, the framework should be reduced or revised. If only a specific proposition fails, that proposition should be revised without treating its failure as sufficient to reject the analytic level as a whole.

## 8. Conclusion

As generative AI becomes a sustained participant in everyday cognitive activity, the important question is no longer only whether AI raises or lowers a particular performance outcome or psychological variable. Existing research already shows that cognitive work can be offloaded, monitoring and control can be redistributed, task representations can change, and human–AI activity can unfold dynamically. A further theoretical task is to analyze more precisely the process differences that arise under sustained AI participation, how those differences are reconfigured through interaction, and how they may change as experience accumulates. The present article proposes a relational-organization level for this task. Rather than describing change only through the levels and covariation of one or more constructs, or treating the human–AI system as the sole unit of analysis, the framework anchors analysis in the person's current task-cognitive state and examines how that human-side state is relationally configured with task-relevant cognitive operations and AI participation across five dimensions: execution locus, cognitive governance, representational reorganization, process organization, and reachable cognitive space. AI may enter as one endpoint of cross-boundary relations, but its internal states are not incorporated into the human-side analytic object. This five-dimensional decomposition is not an exhaustive taxonomy of human cognitive structure, but a testable and revisable theoretical analytic structure. It provides a structured way to characterize process differences and their dynamic change under sustained AI participation when aggregate indicators or construct-level analyses do not uniquely determine those differences.

The five relational dimensions are neither five new psychological constructs nor five mutually exclusive user types. They are analytically distinguishable and dynamically coupled. Actions generated by the current organization, AI responses, outcomes, costs, and experiences may in turn selectively reweight the future probabilities of different organizational pathways. A person's frequency of AI use can therefore remain stable while verification practices, governance relations, and process structures continue to change; immediate outcomes likewise cannot uniquely reveal the cognitive process that produced them or the trajectory that follows.

Over longer time scales, the framework's most cautious but consequential prediction is that if different cognitive organizations recur across repeated or functionally comparable tasks, they should form different distributions of cognitive practice opportunities; these distributions can be tested using independently accumulated cross-task indicators of practice frequency, the proportion undertaken independently, effort investment, or feedback exposure. If these differences accumulate over time, they may further correspond to different developmental trajectories in strategies, habits, and abilities. The framework does not assume that overall cognitive ability will uniformly improve or decline in a single direction. The critical empirical question is which forms of human cognitive organization are repeatedly formed, maintained, adjusted, or abandoned under which task conditions in sustained human–AI cognition.

Future research should therefore not stop at asking how much AI people use, how one or more psychological constructs change, or whether immediate outcomes improve. A more demanding test is whether the relational-organization framework can reveal cognitive process structures that aggregate indicators and construct-level analyses do not uniquely determine: how a human-side relational organization anchored in the person's current task-cognitive state forms different configurations across the five relational dimensions under sustained AI participation, how those configurations change selectively with interaction history, and whether repeated establishment of particular organizations across repeated or functionally comparable tasks systematically redistributes cognitive practice opportunities on independently accumulated cross-task indicators and may subsequently correspond to different trajectories in strategies, habits, and abilities. The framework's theoretical value ultimately depends not on the number of new terms it introduces, but on whether, relative to aggregate use indicators, construct-level analyses, and existing configuration descriptions, it provides additional measurable and testable discriminative and predictive value. If it does not, the framework should be reduced, revised, or reorganized accordingly.